\documentclass[twocolumn,showpacs,superscriptaddress,prl]{revtex4}

\usepackage{amsmath,amssymb}
\usepackage{graphicx}
\usepackage{verbatim}

\usepackage{amsmath}

\draft
\begin{document}

\title{The Scaling of the Anomalous Hall Effect in the Insulating Regime}

\author{Xiong-Jun Liu, Xin Liu}
\affiliation{Department of Physics, Texas A\&M University, College Station, Texas 77843-4242, USA}
\author{Jairo Sinova}
\affiliation{Department of Physics, Texas A\&M University, College Station, Texas 77843-4242, USA}
\affiliation{Institute of Physics ASCR, Cukrovarnická 10, 162 53 Praha 6, Czech
Republic }

\begin{abstract}
We develop a theoretical approach to study the scaling of anomalous Hall effect (AHE) in the insulating regime, which is observed to be $\sigma_{xy}^{AH}\propto\sigma_{xx}^{1.40\sim1.75}$ in experiments over a large range of materials. This scaling is qualitatively different from the ones observed in metals. Basing our theory on the phonon-assisted hopping mechanism and percolation theory, we derive a general formula for the anomalous Hall conductivity, and show that it scales with the longitudinal conductivity as $\sigma_{xy}^{AH}\sim\sigma_{xx}^{\gamma}$ with $\gamma$ predicted to be $1.38\leq\gamma\leq1.76$, quantitatively in agreement with the experimental observations. Our result provides a clearer understanding of the AHE in the insulating regime and completes the scaling phase diagram of the AHE.
\end{abstract}
\date{\today }
\maketitle


%
The anomalous Hall effect (AHE) is a central topic in the study of Ferromagnetic materials \cite{AHE1}.
It exhibits the empirical relation $\rho_{xy}=R_0B_z+R_SM_z$ between the total Hall resistivity and the magnetization $M_z$ and external magnetic field $B_z$.
Here $R_0$ and $R_S$ are respectively the ordinary and anomalous Hall coefficients.
When transformed to an anomalous Hall conductivity (AHC), $\sigma_{xy}^{AH}$, three regimes are observed with respect to its dependence on the diagonal conductivity, $\sigma_{xx}$.
In the metallic regime the AHE $\sigma_{xy}^{AH}$ is observed to be linearly proportional to $\sigma_{xx}$ for the highest metallic systems
($\sigma_{xx}>10^6 \Omega^{-1}$ cm$^{-1}$) and roughly constant for the rest of
the metallic regime. This dependence indicates the different dominant mechanisms in ferromagnetic metals. 
These are understood to be the  skew scattering, side jump scattering, and intrinsic deflection mechanisms.
The intrinsic contribution  is induced by a momentum-space Berry phase linked
to the electronic structure of the multi-band SO coupled system \cite{AHE1,Luttinger}.
The side jump scattering mechanism  gives the same scaling relation as the intrinsic contribution, i.e. $\sigma_{xy}^{AH-sj}\propto\sigma_{xx}^0$,
and the skew scattering is linear in the longitudinal conductivity, $\sigma_{xy}^{AH-sk}\propto\sigma_{xx}$. While these mechanisms are now better understood,
the maximum scaling exponent of the AHC cannot exceed unity in the metallic regime \cite{AHE1}.

On the other hand, experiments in the insulating regime exhibit an unexpected scaling relation of the AHC $\sigma_{xy}^{AH}\propto\sigma_{xx}^{1.40\sim1.75}$,
which remains unexplained and a major challenge in understanding fully the phase diagram of the AHE
\cite{insulating1,insulating2,insulating3,insulating4,insulating5,insulating6,insulating7, insulating8,insulating9,insulating10,insulating11}.
The available microscopic theories of metals fail in this regime since the condition $k_Fl\gg1$  is no longer satisfied for  disordered insulators \cite{AHE1,Onoda}.
The few previous studies of the AHE in the insulating regime focused on manganites and Ga$_{1-x}$Mn$_x$As; while the
manganites do not exhibit this scaling, the studies on insulating Ga$_{1-x}$Mn$_x$As did not show this scaling \cite{Manganites,Burkov,note}.

In this Letter we study the scaling of the AHE in
the insulating strongly disordered amorphous regime, where  at low temperatures
charge transport results from phonon-assisted hopping between impurity localized states \cite{Abrahams,Mott}. We calculate the upper and lower limits of the AHC, and show it scales with $\sigma_{xx}$ as $\sigma_{xy}^{AH}\sim\sigma_{xx}^{\gamma}$ with $\gamma$ predicted to be $1.38\leq\gamma\leq1.76$, in agreement with the experimental observations.
This scaling remains the same regardless of whether the hopping process is Mott-variable-range-hopping or influenced by interactions, i.e. Efros-Shkolvskii (ES) regime.

\begin{figure}[ht]
\includegraphics[width=0.65\columnwidth]{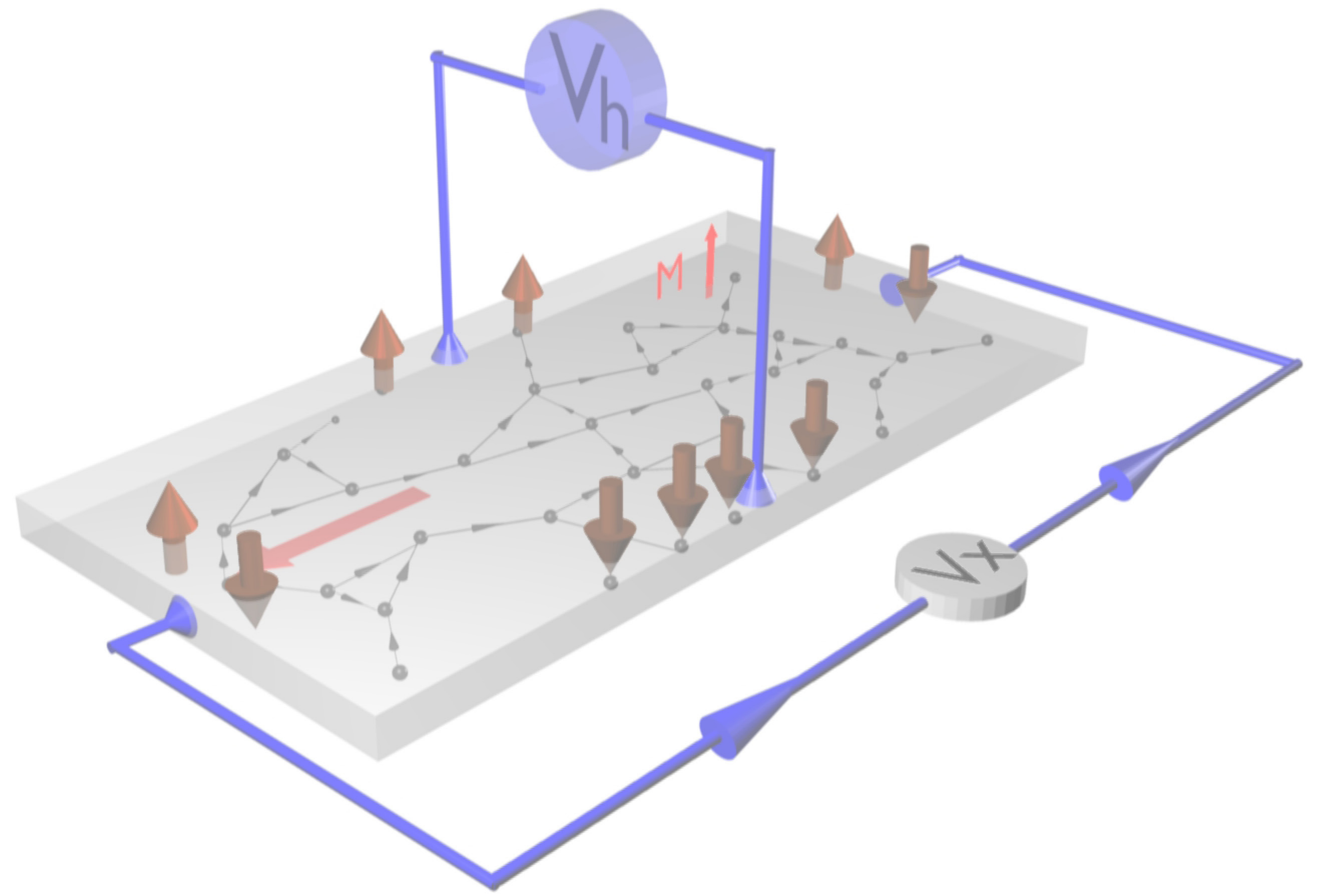}
\caption{AHE in the insulating regime. In this regime charge transport occurs via hopping between impurity sites.}
\label{sketch}
\end{figure}

\begin{figure}[ht]
\includegraphics[width=1.0\columnwidth]{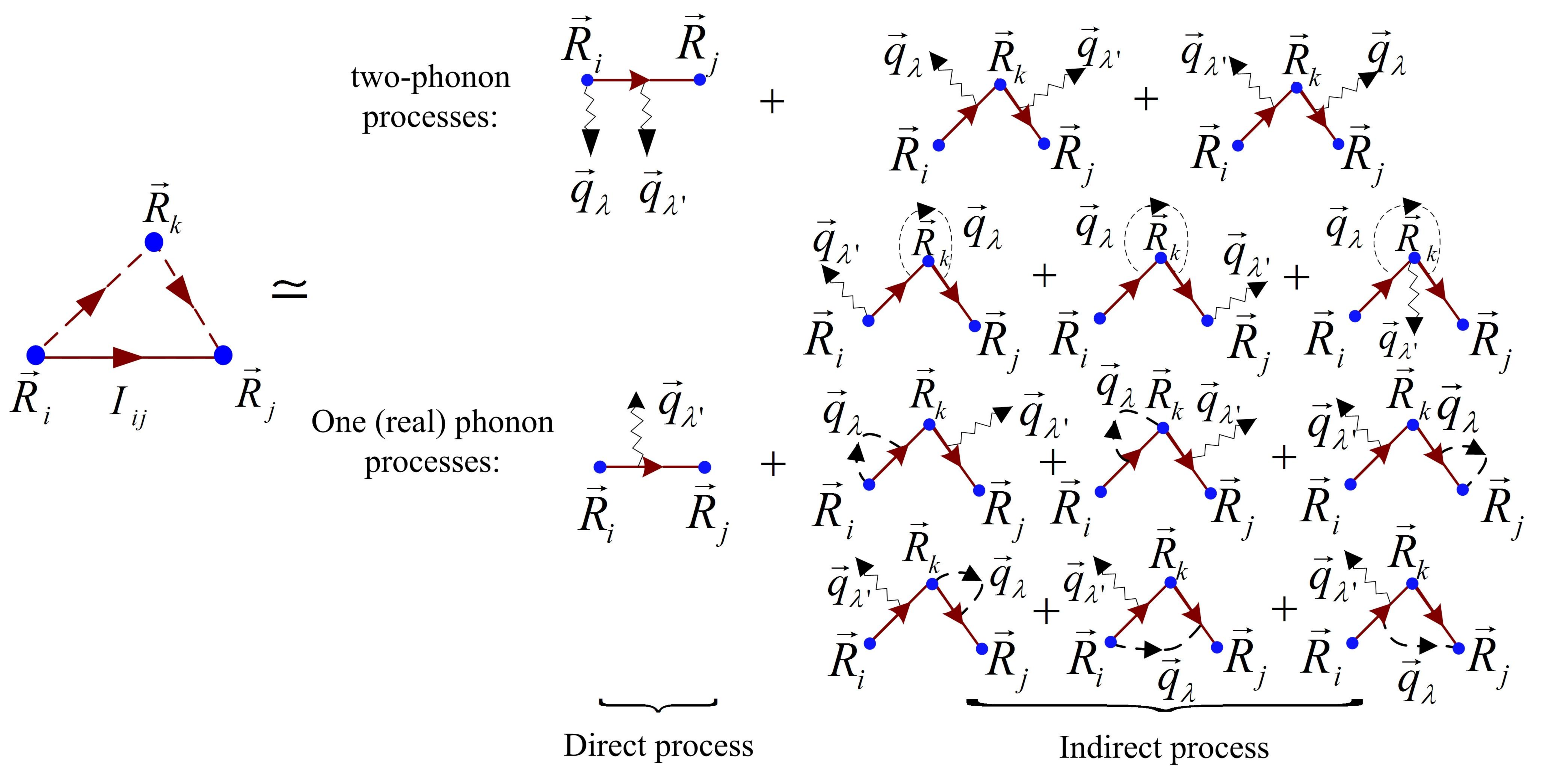}
\caption{(Color online) The hopping processes through triads with up to two real phonons absorbed or emitted. (Top) Typical diagrams of the two-phonon direct and indirect hopping processes. (Bottom) One-phonon direct process and typical three-phonon (one real phonon) indirect hopping processes.}
\label{multi}
\end{figure}

To capture the Hall effect one requires the hopping process between impurity sites (Fig.~\ref{sketch}) to break the time-reversal (TR) symmetry.
The two-site direct hopping preserves TR symmetry, and therefore
more than two sites must be considered.
The hopping through three sites, as depicted in Fig. \ref{multi}, is the minimum requirement to model theoretically the ordinary Hall effect (OHE) \cite{Holstein}.
The total hopping amplitude is obtained by adding  the direct and indirect (through the intermediate $k$-site) hopping terms from $i$ to  $j$ sites.
The two hopping paths give rise to an interference term for the transition rate which breaks TR symmetry and is responsible for the Hall current in the hopping regime. For the OHE, the interference is a reflection of the Aharonov-Bohm phase, and for the AHE it reflects the Berry phase due to SO coupling. Furthermore, the dominant contribution to the Hall transport will be given by the one- and two-real-phonon processes through triads (Fig. \ref{multi}) \cite{Holstein}.



Our theory is based on a minimal tight-binding Hamiltonian. 
With the particle-phonon coupling considered, the total Hamiltonian $H=H_p+H_{c}+H_{ph}$, with
\begin{eqnarray}\label{eqn:threesite1}
H_p&=&\sum_{i\alpha}\epsilon_i\hat c_{i\alpha}^\dag\hat c_{i\alpha}-\sum_{i\alpha,j\beta}t_{i\alpha,j\beta}\hat c_{i\alpha}^\dag\hat c_{j\beta}+\sum_{i\alpha\beta}\bold M\cdot\mathbf{\tau}_{\alpha\beta}\hat c_{i\alpha}^\dag\hat c_{i\beta}\nonumber\\
H_{c}&=&i\eta\sum_{i\alpha\lambda}(\vec q_\lambda\cdot\vec e_\lambda)\omega_\lambda^{-1/2}(b_\lambda e^{i\vec{q}_\lambda\cdot\vec{r}}-b_\lambda^\dag e^{-i\vec{q}_\lambda\cdot\vec{r}})\hat c_{i\alpha}^\dag\hat c_{i\alpha}\nonumber\\
H_{ph}&=&\sum_{\lambda}\omega_\lambda b_\lambda^\dag b_\lambda.\nonumber
\end{eqnarray}
Here $H_p$ describes localized states, $H_{c}$ gives the particle-phonon coupling with $\eta$ the coupling constant, $H_{ph}$ is the phonon Hamiltonian, $\alpha$ is the local on-site total angular momentum index, and $\epsilon_i$ is the energy measured from the fermi level. Here we consider that the magnetization is 
saturated and thus  assume $\bold M=M\hat e_z$. The hopping matrix $t_{ij}$ is generally off-diagonal due to SO coupling (see Supplementary Information (SI)).
The localization regime 
has the condition $|t_{i\alpha,j\beta}|\ll|\epsilon_i-\epsilon_j|$ in average. The specific form of the relevant parameters ($t_{ij}$, $M$, spin operator $\mathbf{\tau}_{\alpha\beta}$)
are material dependent and do not affect the scaling relation between $\sigma_{xy}^{AH}$ and $\sigma_{xx}$.

Considering the dominant contributions to the longitudinal and Hall transports, we obtain the charge current between $i$ and $j$ sites in a single triad with applied voltages \cite{Burkov}:
$I_{ij}=G_{ij}V_{ij}+{\cal G}_{ijk}(V_{ik}+V_{jk})$,
with $G_{ij}$ the direct conductance and ${\cal G}_{ijk}$ responsible for Hall transport.
The formula of $I_{ij}$ gives the microscopic conductances in any single triad (see SI). To evaluate the macroscopic AHC, we need to properly average it over all triads in the random system. This is achieved with the aid of percolation theory, a fundamental tool to understand the hopping transport.

We first map the random impurity system to a random resistor network by introducing the connectivity between impurity sites with the help of a cut-off $G_c(T)$. When the conductance between two impurity sites satisfies $G_{ij}\geq G_c$, we consider the $i,j$ sites are connected with a finite resistor $Z_{ij}=1/G_{ij}$. Otherwise, they are treated as disconnected, i.e. $G_{ij}\rightarrow0$. The Hall effect will be treated as a perturbation to the obtained resistor network. The cut-off $G_c$ should be properly chosen so that the long-range critical percolation paths/clusters appear and span the whole material, and dominate the charge transport in the hopping regime. The macroscopic physical quantities will finally be obtained by averaging over the percolation path/cluter.

The hopping coefficient generally has the form $t_{i\alpha,j\beta}=t_{i\alpha,j\beta}^{(0)}e^{-aR_{ij}}$, with $a^{-1}$ the localization length and $R_{ij}=|\bold R_i-\bold R_j|$. The direct conductance holds the form $G_{ij}=G_0(T)e^{-2aR_{ij}-\frac{1}{2}\beta(|\epsilon_{i\alpha}|+|\epsilon_{j\beta}| +|\epsilon_{i\alpha}-\epsilon_{j\beta}|)}$, and then the cut-off can be introduced by $G_c=G_0e^{-\beta\xi_c(T)}$. Here $\beta\xi_c$ is a decreasing function of $T$, indicating the material in the insulating regime. The number of impurity sites connected to a specific site $i$ with energy $\epsilon_i$ can be calculated by $n(\epsilon_i,\xi_c)=\int d\epsilon_j\int d^3\vec R_{ij} \rho(\epsilon_j,\vec R_{i})\Theta\bigr(G_{ij}-G_c\bigr)$.
Here $\Theta(x)$ is the step function and the DOS $\rho(\epsilon,\vec R_{i})\approx\frac{1}{V}\sum_{i}\delta(\epsilon-\epsilon_i)$ is
approximated to be spatially homogeneous.
The number $n(\epsilon_i,\xi_c)$ can also be given by $n(\epsilon_i,\xi_c)=\sum_nP_n(\epsilon_i,\xi_c)$, with $P_n(\epsilon_i,\xi_c)$ being the probability that the $n$-th smallest resistor connected to the site $i$ has the resistance less than $1/G_c$. The function $P_n$ reads $P_n(\epsilon_i,\xi_c)=\frac{1}{(n-1)!}\int_0^{n(\epsilon_i)}e^{-x}x^{n-1}dx$ \cite{Pollak}.
The percolation path/cluster appears when the average connections per impurity site $\bar{n}=\langle n(\epsilon_i)\rangle_c$
reaches the critical value $\bar{n}_c$, where the definition of
$\langle...\rangle_c$ is given in Eq.~(\ref{eqn:averaging1}).
Suppose a physical quantity $F(\epsilon_1,...,\epsilon_m;\vec r_1,...,\vec r_m)$ being a $m$-site function, requiring  the $i$-th site to have at least $\eta_i$ sites connected to it.
The averaging of $F(\epsilon;\vec r)$ reads
\begin{eqnarray}\label{eqn:averaging1}
\langle F(\epsilon;\vec r)\rangle_c&=&\frac{1}{{\cal N}_F}\int d\epsilon_1...\int d\epsilon_m\int d^3\vec r_{12}...\int d^3\vec r_{m-1,m}\nonumber\\
&&\times\prod_{i=1}^{m}{\cal P}_{\eta_i}(\epsilon_i)F(\epsilon_1,...,\epsilon_m;\vec r_1,...,\vec r_m),
\end{eqnarray}
where ${\cal N}_F$ is a normalization factor and
the probability function ${\cal P}_{\eta_i}(\epsilon_i)=\rho(\epsilon_i)\sum_{k\geq \eta_i}P_{k}(\epsilon_i)$.
The term $\sum_{k\geq \eta_i}P_{k}(\epsilon_i)$ entering the probability function has important physical reason.
The configuration averaging is not conducted over the whole impurity system, but over the percolation cluster which covers only portion of the impurity sites. Therefore the probability that an impurity site belonging to the percolation cluster must be taken into account for probability function. Moreover, this probability function also distinguishes the physical origins of the AHC and $\sigma_{xx}$.
For $\sigma_{xy}^{AH}$ one has $\eta_i=3$, and for $\sigma_{xx}$ one has $\eta_i=2$.
This indicates the averaging of $\sigma_{xx}$ is performed along the one dimensional (1D) percolation path, while for AHE which is a two dimensional (2D) effect, one shall evaluate AHC over all triads connected in the 2D percolation cluster.

\begin{figure}[ht]
\includegraphics[width=0.7\columnwidth]{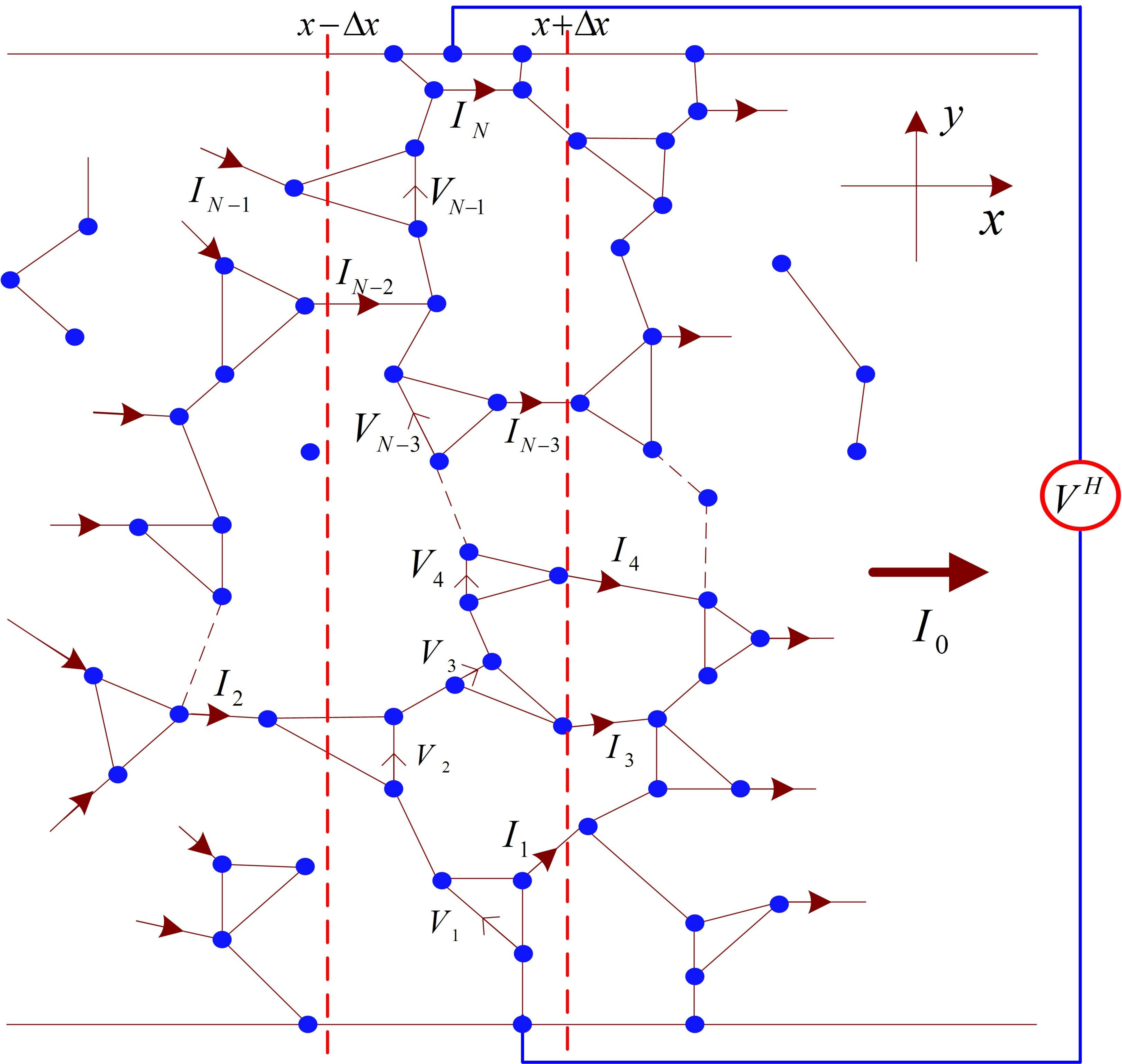}
\caption{(Color online) Typical resistor network in the material. The present situation indicates $V_{N-2}^H$ and $V_{N}^H$ in the region from $x-\Delta x$ to $x+\Delta x$ are zero, where no triads form.}
\label{percolation}
\end{figure}

Numerical solutions show the critical site connectivity is $\bar{n}_c=2.6\sim2.7$ for the appearance of a percolation path/cluster in three dimensional materials \cite{hopping1,hopping2}. This indicates the triads are sparsely distributed in the percolation cluster, as shown in Fig.~\ref{percolation}. The AHC can be derived by examining the transverse voltage $V_y^H$ (along the $y$-axis) induced by the applied longitudinal current $I_0$.
Denote by $N(x)$ the number of triads distributed along the $y$-axis in the region around position $x$
(Note $\bold M$ is along the $z$-axis, hence we assume the system in this direction to be uniform). The transverse voltage equals the summation over the voltage drops of the $N(x)$ triads: $V_y^H(x)=\sum_{l=1}^{N(x)}V_l^H$.
The average Hall voltage $\bar{V}_y^H$ can be obtained in the limit $N(x)\rightarrow\infty$, which from Eq. (\ref{eqn:averaging1}) we find (see SI for details)
\begin{eqnarray}\label{eqn:anomaloushall1}
\sigma_{xy}^{AH}=3L\sigma_{xx}^2\frac{k_BT}{e^2}\langle\frac{\sum_{\alpha\beta\gamma}\bigr[\mbox{Im}(t_{i\alpha,j\beta} t_{j\beta,k\gamma}t_{k\gamma,i\alpha})T_{ijk}^{(3)}\bigr]} {\sum_{i\leftrightarrow j\leftrightarrow k}|t_{ij}t_{jk}|^2T_{ij}^{(2)}T_{jk}^{(2)}}\rangle_c,
\end{eqnarray}
with $L$ the correlation length of the network. Note the configuration integral given by Eq. (\ref{eqn:averaging1}) is first derived for the AHC in this letter. This is an essential difference from the former theory by Burkov {\it et al} \cite{Burkov}, where the configuration averaging applies to the whole system rather than to 2D percolation cluster. With our formalism the key physics that Hall currents are averaged over percolation clusters can be studied, which is a crucial step to understand the insulating regime of the AHE phase diagram.
The above configuration integral cannot be solved analytically. In the following we study the upper and lower limits of the AHC by imposing further restrictions in Eq.~(\ref{eqn:anomaloushall1}), with which the range of the scaling relation between $\sigma_{xy}^{AH}$ and $\sigma_{xx}$ can be determined.

The lower (upper) limit of the AHC can be formulated by keeping only the maximum (minimum) term in the denominator and the minimum (maximum) term in the numerator.
Furthermore, for simplicity 
we first  approximate  the DOS to be constant although this approximation is relaxed later. 
As a result, with further simplification (see SI) we find
\begin{eqnarray}\label{eqn:anomaloushall2}
\{\sigma_{xy}^{AH}\}_{\substack{min\\max}}\simeq3L\sigma_{xx}^2\frac{k_BT}{ e^2t_{\substack{max/min}}^{(0)}}\langle R_{ijk}^{\substack{min\\max}}\rangle_c\langle\epsilon_{ijk}^{\substack{min\\max}}\rangle_c,
\end{eqnarray}
where $\langle R_{ijk}^{min}\rangle_c=e^{a\langle R_{ij}+R_{jk}-R_{ik}\rangle_c}|_{R_{ij},R_{jk}<R_{ik}}, \langle\epsilon_{ijk}^{min}\rangle_c=e^{0.5\beta\langle|\epsilon_{i}|+|\epsilon_{j}|+ |\epsilon_{j}-\epsilon_{k}|-|\epsilon_{i}-\epsilon_{k}|\rangle_c}|_{|\epsilon_{i}|<|\epsilon_{j}| <|\epsilon_{k}|}$, $\langle R_{ijk}^{max}\rangle_c$ and $\langle\epsilon_{ijk}^{max}\rangle_c$ hold the same form for the calculation but the restrictions change to be $R_{ij},R_{jk}>R_{ik}$ and $|\epsilon_{i}|>|\epsilon_{j}|>|\epsilon_{k}|$, respectively. The coefficient $t_{\rm max/min}^{(0)}$ represents the maxmimum/minimum element in the matrix $t_{ij}^{(0)}$.
It is instructive to point out the underlying physics of the two limits.
In the hopping regime, charge transport may prefer a short and straight path in the forward direction with larger resistance than a long and meandrous path with somewhat smaller resistance \cite{Abrahams,Pollak}. This picture introduces an additional restriction complementary to the percolation theory for charge transport. What bonds in a triad play the major role for the current flowing through it is determined by the optimization of the resistance magnitudes and spatial configuration of the three bonds. A quantitative description can be obtained by phenomenologically introducing an additional probability factor to restrict the charge transport  \cite{Abrahams,Pollak}.
Here we only need to adopt this picture to present the two extreme situations corresponding to $\{\sigma_{xy}^{AH}\}_{min/max}$.
To get the upper limit we assume that for each triad of the percolation cluster the two bonds with smaller direct conductance dominate the charge transport, i.e.
the product of two smallest conductances minimize the denominator, and take the maximum value for the numerator of Eq. (\ref{eqn:anomaloushall1}).
The opposite limit corresponds to the situation that the two bonds with larger conductances in each triad dominate the charge transport.

For a constant DOS, one obtains straightforwardly the number $n(\epsilon_i)$ and then the probability $P_n(\epsilon_i)$. Substituting them into Eq.~(\ref{eqn:anomaloushall2}) we finally obtain $\langle R_{ijk}^{min}\rangle_c\simeq e^{0.156\beta\xi_c},
\langle R_{ijk}^{max}\rangle_c\simeq e^{0.483\beta\xi_c}$,
$\langle\epsilon_{ijk}^{min}\rangle_c\simeq e^{0.086\beta\xi_c}$, and
$\langle\epsilon_{ijk}^{max}\rangle_c\simeq e^{0.138\beta\xi_c}$ (see SI for details).
The longitudinal conductivity is obtained based on the 2-site function $G_{ij}$ which should be no less than $G_c$ in a percolation path. The result of $\sigma_{xx}$ equals $G_c$ divided by the correlation length of the network and takes the form $\sigma_{xx}=\sigma_0(T)e^{-\beta\xi_c}$, where $\sigma_0(T)$ gives at most a power-law on $T$ \cite{Pollak,Halperin}. Comparing this form with the AHC, we reach
$\{\sigma_{xy}^{AH}\}_{min/max}\sim\sigma_{0}^{2-\gamma_{a/b}}\sigma_{xx}^{\gamma_{a/b}}$ with $\gamma_{a}=1.76$ and $\gamma_{b}=1.38$.
This leads to the scaling relation, the central result of this Letter, between $\sigma_{xy}^{AH}$ and $\sigma_{xx}$
 of the AHE in the insulating regime:
\begin{eqnarray}\label{eqn:maximinimum1}
\sigma_{xy}^{AH}\propto\sigma_{xx}^{\gamma}, \ \ 1.38<\gamma<1.76.
\end{eqnarray}
The maximum (minimum) of the AHC corresponds to the smaller (larger) power index $\gamma_b$ ($\gamma_a$).
This scaling range can be confirmed with a numerical calculation of the Eq. (\ref{eqn:anomaloushall2}). Furthermore, a direct numerical study for the configuration integral (\ref{eqn:anomaloushall1}) gives the scaling exponent $\gamma\approx1.62$, which is consistent with our prediction of the lower and upper limits.


So far in the calculation we have assumed a constant DOS. This approximation is applicable for the
ferromagnetic system with strong exchange interaction between local magnetic moments and charge carriers
(e.g. oxides, magnetites) and 
half metals in general. In this case we do not need to sum over spin-up and spin-down states which contribute oppositely to the AHE, and the previous results are valid.

However, when the Fermi energy crosses both spin-up and -down impurity states, a symmetric DOS with $\rho(\epsilon)=\rho(-\epsilon)$ leads to zero AHC.
This is because under the transformation $\epsilon_{l,\sigma}\rightarrow-\epsilon_{l,-\sigma}$ ($l=i,j,k$), ${\cal G}_{ijk}$ changes sign, while $G_{ij}$ is invariant.
Thus the averaging for AHC over all spin states and on-site energies cancels \cite{Burkov}.
We relax the previous simplifying restriction by  expanding the DOS by $\rho(\epsilon)=\sum_n\frac{1}{n!}\frac{d^n\rho_0}{d\epsilon^n_F}\epsilon^n$,
where $|\epsilon|\leq\xi_c$ and we consider $\rho_0=\rho(\epsilon_F)>0$. Substituting this expansion into Eq.~(\ref{eqn:anomaloushall1}) yields $\sigma_{xy}^{AH}=\sum_{n=0}^{\infty}\sigma_{xy}^{(n)}$, with the $1$st and $2$nd nonzero terms respectively proportional to $\frac{d\rho_0}{d\epsilon_F}$ and $\frac{d^3\rho_0}{d\epsilon_F^3}$.
We can similarly evaluate the lower and upper limits of $\sigma_{xy}^{AH}$ as before. The first two nonzero terms in the expansion are $\{\sigma_{xy}^{(1)}\}_{min/max}\sim M\frac{d\rho_0}{d\epsilon_F} \xi_c(T)\sigma_{0}^{2-\gamma_{a/b}}\sigma_{xx}^{\gamma_{a/b}}$ and $\{\sigma_{xy}^{(2)}\}_{min/max}\sim0.002M\frac{d^3\rho_0}{d\epsilon_F^3} \xi_c^3(T)\sigma_{0}^{2-\gamma_{a/b}}\sigma_{xx}^{\gamma_{a/b}}$.
The appearance of $M$ is due to the summation over the spin-up and -down states. We have also employed the result $\langle |\epsilon|\rangle_c=0.112\xi_c$. The specific formulas of $\sigma_0(T)$ and $\xi_c(T)$ do not affect the qualitative scaling between $\sigma_{xy}^{AH}$ and $\sigma_{xx}$. For the Mott and ES hopping regimes, we have respectively $\xi_c=k_BT\bigr(T_0/T\bigr)^{1/4}$ and $\xi_c=k_BT\bigr(T_0/T\bigr)^{1/2}$ with $T_0$ the constant depending on the DOS \cite{Pollak,Halperin,Efros}. Note that $\sigma_{xy}^{(1)}$ and $\sigma_{xy}^{(2)}$ have different physical meanings. The term $\sigma_{xy}^{(1)}$ dominates when the DOS varies monotonically versus $\epsilon$. Furthermore, when the DOS has a local minimum at the Fermi level, which may be obtained due to particle-particle interaction (coulomb interaction), we have $d\rho/d\epsilon_F=0$. Then the term $\sigma_{xy}^{(1)}$ varnishes and $\sigma_{xy}^{(2)}$ dominates the AHE. The above results also indicate that the AHC may change sign when $d\rho_0/d\epsilon_F$ or $d^3\rho_0/d\epsilon_F^3$ changes sign, which is consistent with the observation by Allen {\it et al} \cite{insulating4}.
\begin{figure}[ht]
\includegraphics[width=0.8\columnwidth]{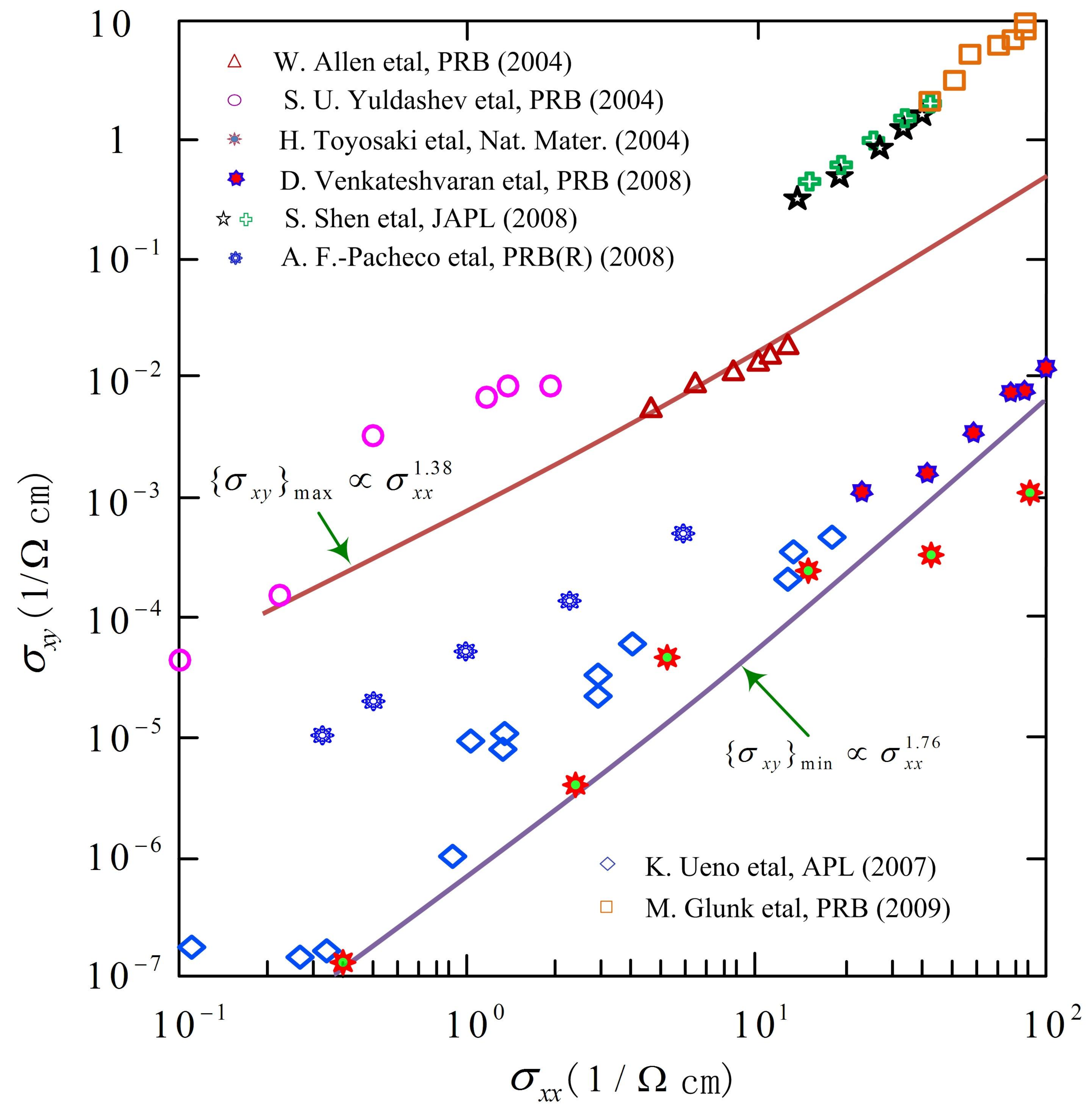}
\caption{(Color online) Scaling relation between the AHC and longitudinal conductivity. The theoretical results are compared with the experimental observations.}
\label{scaling}
\end{figure}


Fig. \ref{scaling} shows our theoretical prediction is consistent with the experimental observations of the scaling relation in this regime, hence completing the understanding of the phase diagram of the
AHE.

This work is supported by  NSF under Grant No. DMR-0547875, NSF-MRSEC DMR-0820414, NHARP, and by SWAN-NRI, and the
 Research Corporation for the Advancement of Science.


\noindent

\begin{widetext}
\section{Supplementary Information for ``Scaling of the Anomalous Hall Effect in the Insulating Regime"}

\section{Hopping matrix}

In the case the magnetization is saturated and thus $\bold M=M\hat e_z$, we rewrite the Hamiltonian $H_p$ in the diagonal basis of the exchange term and obtain
\begin{eqnarray}\label{eqn:threesite2}
H_p=\sum_{\alpha}\epsilon_{i\alpha}\hat c_{i\alpha}^\dag\hat c_{i\alpha}-\sum_{i\alpha,j\beta}t_{i\alpha,j\beta}\hat c_{i\alpha}^\dag\hat c_{j\beta},
\end{eqnarray}
where $\epsilon_{i\alpha}=\epsilon_i+M\tau_{\alpha\alpha}$. Below are two different examples. First, for the dilute Ga$_{1-x}$Mn$_x$As, the matrix $t_{i\alpha,j\beta}$ describes the hopping of the holes localized on the Mn impurities. Under the spherical approximation $t_{i\alpha,j\beta}$ can be obtained based on by a unitary rotation $U(\bold R_{ij})$ from the $\hat e_z$ direction to the hopping direction $i\rightarrow j$ \cite{DMS}. We thus have $t_{i\alpha,j\beta}=[U^\dag(\bold R_{ij})t_{diag}U(\bold R_{ij})]_{\alpha\beta}$ with $t_{diag}=\mbox{diag}[t_{3/2},t_{1/2},t_{-1/2},t_{-3/2}]$ representing the situation that the hopping direction is along the $z$ axis. Another case is for the localized $s$-orbital electrons. In this case, the hopping is given by $t_{ij}=U^\dag(\bold R_{ij})[\tilde{t}_{ij}(1+i\vec{v}_{ij}\cdot\vec\sigma)]U(\bold R_{ij})$. Here $\tilde{t}_{ij}=\mbox{diag}[t_{1/2},t_{-1/2}]$ and $\vec{v}_{ij}=\frac{\alpha}{\hbar}\int_{\vec{r}_i}^{\vec{r}_j}(\nabla V(\bold r)\times d\vec{r}')$ with $V(\bold r)$ including the ion and external potentials, the spin-orbit coupling coefficient $\alpha=\hbar/(4m^2c^2)$ and $m$ the effective mass of the electron.

\section{Configurational integrals}

The averaging of a $m$-site physical quantity $F(\epsilon_1,\epsilon_2,...,\epsilon_m;\vec r_1,\vec r_2,...,\vec r_m)$ along critical percolation path/cluster is given by
\begin{eqnarray}\label{eqn:averaging1}
\langle F(\epsilon_1,\epsilon_2,...,\epsilon_m;\vec r_1,\vec r_2,...,\vec r_m)\rangle_c&=&\frac{1}{{\cal N}_F}\int d\epsilon_1\int d\epsilon_2...\int d\epsilon_m\times\nonumber\\
&\times&\int d^3\vec r_{12}\int d^3\vec r_{23}...\int d^3\vec r_{m-1,m}\rho(\epsilon_1)\sum_{k=n_1}^{\infty}P_{k}(\epsilon_1)\rho(\epsilon_2)\times\nonumber\\
&\times&\sum_{k=n_2}^{\infty}P_{k}(\epsilon_2)... \rho(\epsilon_m)\sum_{k=n_m}^{\infty}P_{k}(\epsilon_m)F(\epsilon_1,\epsilon_2,...,\epsilon_m;\vec r_1,\vec r_2,...,\vec r_m),
\end{eqnarray}
where $P_n(\epsilon_i,\xi_c)=\frac{1}{(n-1)!}\int_0^{n(\epsilon_i)}e^{-x}x^{n-1}dx$ \cite{Pollak}. Some examples are given below. The first one is the average value of $n(\epsilon_i,\xi_c)$ in the percolation cluster. Note $n(\epsilon_i,\xi_c)$ is a $1$-site function. The averaging is straightforward and
\begin{eqnarray}\label{eqn:averaging3}
\bar{n}=\langle n(\epsilon,\xi_c)\rangle_c=\frac{\int d\epsilon_in(\epsilon_i)\rho(\epsilon_i)\sum_{k=1}^{\infty}P_{k}(\epsilon_i)}
{\int d\epsilon_\rho(\epsilon_i)\sum_{k=1}^{\infty}P_{k}(\epsilon_i)}=\frac{\int d\epsilon_in(\epsilon_i)\rho(\epsilon_i)n(\epsilon_i)}
{\int d\epsilon_in(\epsilon_i)\rho(\epsilon_i)}.
\end{eqnarray}
The hopping conduction occurs when the average value $\bar{n}$ reaches the critical value $\bar n_c$. When the DOS $\rho(\epsilon_i)=\rho_0$ is a constant, the number $n(\epsilon_i)$ is given by $n(\epsilon_i)=\frac{2\pi}{3}\frac{\rho_0}{(2ak_BT)^3}(\xi_c-|\epsilon_i|)^2(\xi_c^2-|\epsilon_i|^2)$. Then we have
\begin{eqnarray}\label{eqn:connectivity5}
\bar n_c=\frac{2\pi}{3}\frac{\rho_0}{(2ak_BT)^3}\frac{\int(\xi_c-|\epsilon_i|)^4(\xi_c^2-|\epsilon_i|^2)^2d\epsilon_i} {\int(\xi_c-|\epsilon_i|)^2(\xi_c^2-|\epsilon_i|^2)d\epsilon_i}=0.406\pi\frac{\rho_0}{(2ak_BT)^3}\xi_c^4,
\end{eqnarray}
from which we obtain the cut-off value $\xi_c$ by
\begin{eqnarray}\label{eqn:connectivity6}
\xi_c(T)=\biggr[\frac{(2ak_BT)^3\bar n_c}{0.406\pi\rho_0}\biggr]^{1/4}.
\end{eqnarray}
Thus it gives
\begin{eqnarray}\label{eqn:connectivity7}
\beta\xi_c=\biggr(\frac{T_0}{T}\biggr)^{1/4}, \ \ T_0=16\frac{a^3\bar n_c}{k_B\rho_0},
\end{eqnarray}
which is the Mott law. Accordingly, if we assume the density of states $\rho(\epsilon)\sim\epsilon^2$, we obtain straightforwardly the Efros-Shklovskii (E-S) law $\beta\xi_c=\bigr(\frac{T_0}{T}\bigr)^{1/2}$ \cite{E-S}.
Second, we give the formula for the longitudinal resistance based on the $2$-site function $Z_{ij}=1/G_{ij}$. The longitudinal resistance for a percolation path is calculated by
\begin{eqnarray}\label{eqn:averaging4}
\bar{R}_{xx}=\frac{N\int d\epsilon_i\int d\epsilon_j\int d^3\vec r_{ij}Z_{ij}(\epsilon_i,\epsilon_j;\vec r_{ij})\rho(\epsilon_i)\sum_{k=2}^{\infty}P_{k}(\epsilon_i)\rho(\epsilon_j)\sum_{k=2}^{\infty}P_{k}(\epsilon_j)}
{\int d\epsilon_i\int d\epsilon_j\int d^3\vec r_{ij}\rho(\epsilon_i)\sum_{k=2}^{\infty}P_{k}(\epsilon_i)\rho(\epsilon_j)\sum_{k=2}^{\infty}P_{k}(\epsilon_j)},
\end{eqnarray}
where $N$ is the number of links along the percolation path. The above formula can be simplified by the fact that $\sum_{k=2}^{\infty}P_{k}(\epsilon_i)=n(\epsilon_i,\xi_c)-P_1(\epsilon_i,\xi_c)\propto[n(\epsilon_i,\xi_c)]^2$. We then reach
\begin{eqnarray}\label{eqn:averaging5}
\bar{R}_{xx}=\frac{N\int d\epsilon_i\int d\epsilon_j\int d^3\vec r_{ij}Z_{ij}(\epsilon_i,\epsilon_j;\vec r_{ij})\rho(\epsilon_i)[n(\epsilon_i,\xi_c)]^2\rho(\epsilon_j)[n(\epsilon_j,\xi_c)]^2}
{\int d\epsilon_i\int d\epsilon_j\int d^3\vec r_{ij}\rho(\epsilon_i)[n(\epsilon_i,\xi_c)]^2\rho(\epsilon_j)[n(\epsilon_j,\xi_c)]^2}.
\end{eqnarray}
The longitudinal resistivity is given by $\bar{R}_{xx}/(n_dL_x)$, with $n_d$ the density of the percolation paths and $L_x$ the length of the material along $x$ direction \cite{Pollak}. Finally, if the physical quantity is a function of a triad with each site of the triad having at least three sites connected to it, the averaging of such physical quantity is given by
\begin{eqnarray}\label{eqn:averaging6}
\bar{F}(\epsilon_1,\epsilon_2,\epsilon_3;\vec r_1,\vec r_2,\vec r_3)=\frac{\int d\epsilon_1d\epsilon_2d\epsilon_3\int d^3\vec r_{12}\int d^3\vec r_{23}F(\epsilon;\vec r)\rho(\epsilon_1)[n(\epsilon_1)]^3\rho(\epsilon_2)[n(\epsilon_2)]^3 \rho(\epsilon_3)[n(\epsilon_3)]^3}
{\int d\epsilon_1d\epsilon_2d\epsilon_3\int d^3\vec r_{12}\int d^3\vec r_{23}\rho(\epsilon_1)[n(\epsilon_1)]^3\rho(\epsilon_2)[n(\epsilon_2)]^3 \rho(\epsilon_3)[n(\epsilon_3)]^3}.
\end{eqnarray}
The anomalous Hall conductivity/resistivity will be calculated with this formula.

\section{Formula for macroscopic anomalous Hall conductivity}

Now we show rigorously the formula for macroscopic AHC in the hopping regime. The transverse voltage difference for the region from $x-\Delta x$ to $x+\Delta x$ (Fig. 3 in the manuscript) reads
\begin{eqnarray}\label{eqn:hallvoltage1}
V_y(x)=V_1^H+V_2^H+...+V_N^H.
\end{eqnarray}
For the general situation we allow some $V_i^H$'s to be zero (see Fig. 3 in the manuscript). In that case no triad forms for the incoming current $I_i$ under the condition all direct conductances in a triad must be no less than $G_c$.
To calculate $V_i^H$, the voltage contributed by the $i$-th triad, we employ perturbation theory to the equation \cite{perturbation} $I_{ij}=G_{ij}V_{ij}+\sum_{k}{\cal G}_{ijk}(V_{ik}+V_{jk})$. First, in the zeroth order, we consider only the normal current, namely, the Hall current is zero and thus $\sum_jI_{ij}=\sum_jG_{ij}V_{ij}^{(0)}=0$, with which one can determine the voltage $V_{i}^{(0)}$ at each site.
Then, for the first-order perturbation, we have $\sum_jI_{ij}=\sum_jG_{ij}V_{ij}+\sum_jJ_{ij}^{(H)}=0$,
which leads to $J_i^{(H)}=\sum_jJ_{ij}^{(H)}=\sum_j\sum_k{\cal G}_{ijk}(V_{jk}^{(0)}+V_{ik}^{(0)})=-\sum_jG_{ij}V_{ij}$. The current $J_i^{(H)}$ can also be written as
\begin{eqnarray}\label{eqn:hallvoltage4}
J_i^{(H)}=\sum_j\sum_k{\cal G}_{ijk}(V_{jk}^{(0)}+V_{ik}^{(0)})=\frac{1}{2}\sum_{jk}{\cal G}_{ijk}(V_{ik}^{(0)}+V_{jk}^{(0)}-V_{ij}^{(0)}-V_{kj}^{(0)})=\frac{3}{2}\sum_{jk}{\cal G}_{ijk}V_{jk}^{(0)}.
\end{eqnarray}
\begin{figure}[ht]
\includegraphics[width=0.45\columnwidth]{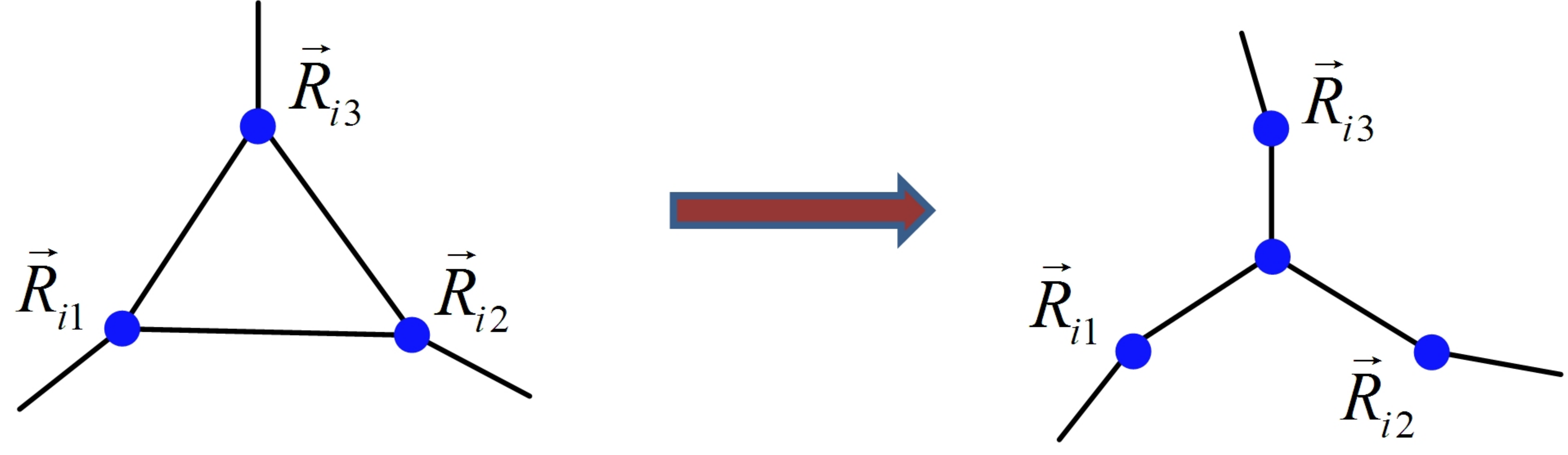}
\caption{(Color online) Resistor network transformation.}
\label{transformation}
\end{figure}
For the hopping regime, the triads are dilutedly distributed and the Hall voltages induced by different triads are considered to be uncorrelated. Therefore, we obtain the Hall voltage of the $i$-th triad from the transformation indicated in fig. \ref{transformation} that
\begin{eqnarray}\label{eqn:hallvoltage5}
V_i^{(H)}=V_{i_3i_2}^{(H)}=\frac{G_{i_1i_2}J_3^{(H)}-G_{i_1i_3}J_2^{(H)}} {G_{i_1i_2}G_{i_2i_3}+G_{i_1i_3}G_{i_2i_3}+G_{i_3i_1}G_{i_1i_2}}=\frac{3I_i{\cal G}_{i_1i_2i_3}^{(i)}} {G_{i_1i_2}G_{i_2i_3}+G_{i_1i_3}G_{i_2i_3}+G_{i_3i_1}G_{i_1i_2}}.
\end{eqnarray}
From the resistor network configuration one can see $\sum_i^{N(x)}I_i=2I_0$. For convenience, we denote $I_i=2I_0\lambda_i(x)$ with $\sum_i\lambda_i=1$. Generally $V_y^H(x)=\sum_iV_i^{(H)}$ is a function of position $x$, and one needs to average it along the $x$ direction.
For a macroscopic system, one has $N(x)\rightarrow\infty$. Furthermore, we consider at the position $x$, for each $\lambda_i$ there are $n_i(x)$ number triads that have such same current fraction $\lambda_i$. Thus we have
\begin{eqnarray}\label{eqn:hallvoltage8}
\bar{V}_y^H=6I_0\frac{1}{L_x}\int dx\sum_{\{n_i\}}\lambda_i\sum_{j=1}^{n_i\gg1}\frac{{\cal G}_{j_1j_2j_3}^{(j)}} {G_{j_1j_2}G_{j_2j_3}+G_{j_1j_3}G_{j_2j_3}+G_{j_3j_1}G_{j_1j_2}},
\end{eqnarray}
To simplify this formula we extend the current distribution $\{\lambda_i\}$ for the region between $x-\Delta x$ and $x+\Delta x$ to the whole space along $x$ direction, and then we can exchange the order of the integral and the first summation: $\frac{1}{L_x}\int dx\sum_{\{n_i\}}\lambda_i\sum_{j=1}^{n_i\gg1}\rightarrow\sum_{\{\lambda_i\}}\lambda_i\frac{1}{L_x}\int dx\sum_{j=1}^{n_i(x)}$. In the limit $N(x)\rightarrow\infty$ and the length $L_x$ much larger than the typical length $L$ of the triad, the calculation $\frac{1}{L_x}\int dx\sum_{j=1}^{n_i(x)}$ gives the average of all possible configurations of the triads through the percolating cluster. This leads to
\begin{eqnarray}\label{eqn:hallvoltage10}
\bar{V}_y^H=6I_0\sum_{\{\lambda_i\}}\bar{n}_i\lambda_i\langle\frac{{\cal G}_{i_1i_2i_3}^{(i)}} {G_{i_1i_2}G_{i_2i_3}+G_{i_1i_3}G_{i_2i_3}+G_{i_3i_1}G_{i_1i_2}}\rangle_c,
\end{eqnarray}
with $\bar{n}_i=(1/L_x)\int dxn_i(x)$ the average number of triads with in/outgoing current $I_i$. Note the identity $\sum_in_i\lambda_i=1$ is independent of position $x$, and therefore we have also $\sum_i\bar{n}_i\lambda_i=1$.
The transverse electric field is given by $\bar{E}_y^H=\bar{V}_y^H/L_y$.
The longitudinal current density reads $j_0=I_0/(L_yL)$, where $L_yL$ represents the area of the cross section. With these results we obtain the Hall conductivity
\begin{eqnarray}\label{eqn:conductivity1}
\sigma_{xy}^{AH}&=&6L\sigma_{xx}^2\langle\frac{{\cal G}_{i_1i_2i_3}} {G_{i_1i_2}G_{i_2i_3}+G_{i_1i_3}G_{i_2i_3}+G_{i_3i_1}G_{i_1i_2}}\rangle_c,\nonumber\\
&=&3L\sigma_{xx}^2\frac{k_BT}{e^2}\langle\frac{\sum_{\alpha\beta\gamma}\bigr[\mbox{Im}(t_{i\alpha,j\beta} t_{j\beta,k\gamma}t_{k\gamma,i\alpha})T_{ijk}^{(3)}\bigr]} {|t_{ij}t_{jk}|^2T_{ij}^{(2)}T_{jk}^{(2)}+|t_{ik}t_{jk}|^2T_{ik}^{(2)}T_{jk}^{(2)}+|t_{ij}t_{ik}|^2T_{ij}^{(2)}T_{ik}^{(2)}}\rangle_c,
\end{eqnarray}
where $T_{jk}^{(2)}$ and $T_{ijk}^{(3)}$ are defined by
\begin{eqnarray}\label{eqn:T1}
T_{ij}^{(2)}=|\Delta_{ij}|e^{-\frac{1}{2k_BT}(|\epsilon_{i\alpha}|+|\epsilon_{j\beta}|+|\epsilon_{i\alpha}-\epsilon_{j\beta}|)},
\end{eqnarray}
with $\Delta_{ij}=\epsilon_{i\alpha}-\epsilon_{j\beta}$, and
\begin{eqnarray}\label{eqn:T2}
T_{ijk}^{(3)}&=&|\Delta_{ij}\Delta_{ik}|e^{-\frac{1}{2k_BT}(|\epsilon_{j\beta}| +|\epsilon_{k\gamma}| +|\epsilon_{i\alpha}-\epsilon_{k\gamma}|+|\epsilon_{i\alpha}-\epsilon_{j\beta}|)}\nonumber\\
&&+|\Delta_{ij}\Delta_{jk}|e^{-\frac{1}{2k_BT}(|\epsilon_{i\alpha}|+|\epsilon_{k\gamma}| +|\epsilon_{j\beta}-\epsilon_{k\gamma}|+|\epsilon_{i\alpha}-\epsilon_{j\beta}|)}\nonumber\\
&&+|\Delta_{ik}\Delta_{kj}|e^{-\frac{1}{2k_BT}(|\epsilon_{i\alpha}|+|\epsilon_{j\beta}|+ |\epsilon_{i\alpha}-\epsilon_{k\gamma}|+|\epsilon_{k\gamma}-\epsilon_{j\beta}|)}.
\end{eqnarray}
The configuration integral will be performed according to the Eq. (\ref{eqn:averaging6}).

\section{Upper and lower limits}

For the lower limit, we let $R_{ij},R_{jk}<R_{ik}$, and $|\epsilon_{i\alpha}|<|\epsilon_{j\beta}|<|\epsilon_{k\gamma}|$. By keeping only the maximum term in the denominator and the minimum one in the numerator of the Eq. (\ref{eqn:conductivity1}) we obtain
\begin{eqnarray}\label{eqn:minimum1}
\{\sigma_{xy}^{AH}\}_{min}&=&3L\sigma_{xx}^2\frac{k_BT}{e^2}\frac{1}{t_{max}^{(0)}}\langle e^{a(R_{ij}+R_{jk}-R_{ik})}e^{\frac{1}{2k_BT}(|\epsilon_{i\alpha}|+|\epsilon_{j\beta}| +|\epsilon_{j\beta}-\epsilon_{k\gamma}|- |\epsilon_{i\alpha}-\epsilon_{k\gamma}|)}\rangle_c\nonumber\\
&\simeq&3L\sigma_{xx}^2\frac{k_BT}{e^2}\frac{1}{t_{max}^{(0)}}\langle e^{a(R_{ij}+R_{jk}-R_{ik})}\rangle_c\langle e^{\frac{1}{2k_BT}(|\epsilon_{i\alpha}|+|\epsilon_{j\beta}| +|\epsilon_{j\beta}-\epsilon_{k\gamma}|- |\epsilon_{i\alpha}-\epsilon_{k\gamma}|)}\rangle_c.
\end{eqnarray}
To make the calculation realistic, we further consider the approximation by replacing the configuration integral of the exponential functions by configuration integral of the exponents. Then we get
\begin{eqnarray}\label{eqn:minimum2}
\{\sigma_{xy}^{AH}\}_{min}\simeq3L\sigma_{xx}^2\frac{k_BT}{e^2}\frac{1}{t_{max}^{(0)}} e^{\langle a(R_{ij}+R_{jk}-R_{ik})\rangle_c|_{R_{ij},R_{jk}<R_{ik}}} e^{\langle\frac{1}{2k_BT}(|\epsilon_{i\alpha}|+|\epsilon_{j\beta}| +|\epsilon_{j\beta}-\epsilon_{k\gamma}|- |\epsilon_{i\alpha}-\epsilon_{k\gamma}|)\rangle_c|_{|\epsilon_{i\alpha}|<|\epsilon_{j\beta}|<|\epsilon_{k\gamma}|}}.
\end{eqnarray}

Similarly, the upper limit can be formulated with the restrictions $R_{ij},R_{jk}>R_{ik}$ and $|\epsilon_{i\alpha}|>|\epsilon_{j\beta}|>|\epsilon_{k\gamma}|$. By the same procedure we obtain
\begin{eqnarray}\label{eqn:minimum1}
\{\sigma_{xy}^{AH}\}_{max}\simeq3L\sigma_{xx}^2\frac{k_BT}{e^2}\frac{1}{t_{min}^{(0)}} e^{\langle a(R_{ij}+R_{jk}-R_{ik})\rangle_c|_{R_{ij},R_{jk}>R_{ik}}} e^{\langle\frac{1}{2k_BT}(|\epsilon_{i\alpha}|+|\epsilon_{j\beta}| +|\epsilon_{j\beta}-\epsilon_{k\gamma}|- |\epsilon_{i\alpha}-\epsilon_{k\gamma}|)\rangle_c|_{|\epsilon_{i\alpha}|>|\epsilon_{j\beta}|>|\epsilon_{k\gamma}|}}.
\end{eqnarray}

\subsection{Lower limit}

First we calculate the lower limit of AHC, which is given by
\begin{eqnarray}\label{eqn:anomaloushall2}
\{\sigma_{xy}^{AH}\}_{\substack{min}}\simeq3L\sigma_{xx}^2\frac{k_BT}{ e^2t_{\substack{max}}^{(0)}}\langle R_{ijk}^{\substack{min}}\rangle_c\langle\epsilon_{ijk}^{\substack{min}}\rangle_c,
\end{eqnarray}
where $\langle R_{ijk}^{min}\rangle_c=e^{a\langle R_{ij}+R_{jk}-R_{ik}\rangle_c}|_{R_{ij},R_{jk}<R_{ik}}, \langle\epsilon_{ijk}^{min}\rangle_c=e^{0.5\beta\langle|\epsilon_{i}|+|\epsilon_{j}|+ |\epsilon_{j}-\epsilon_{k}|-|\epsilon_{i}-\epsilon_{k}|\rangle_c}|_{|\epsilon_{i}|<|\epsilon_{j}| <|\epsilon_{k}|}$. We neglect the spin indices. The configuration integral $\langle R_{ij}+R_{jk}-R_{ik}\rangle_c|_{R_{ij},R_{jk}<R_{ik}}$ is given by
\begin{eqnarray}\label{eqn:position1}
\langle R_{ij}+R_{jk}-R_{ik}\rangle_c=\frac{\int d\epsilon_id\epsilon_jd\epsilon_k\int d^3\bold R_{ij}\int d^3\bold R_{jk}\rho(\epsilon_i)[n(\epsilon_i)]^3\rho(\epsilon_j)[n(\epsilon_j)]^3 \rho(\epsilon_k)[n(\epsilon_k)]^3
(R_{ij}+R_{jk}-R_{ik})}{\int d\epsilon_id\epsilon_jd\epsilon_k\int d^3\bold R_{ij}\int d^3\bold R_{jk}\rho(\epsilon_i)[n(\epsilon_i)]^3\rho(\epsilon_j)[n(\epsilon_j)]^3 \rho(\epsilon_k)[n(\epsilon_k)]^3},
\end{eqnarray}
with $R_{ij},R_{jk}<R_{ik}$. We shall first perform the integral over position $\int d^3\bold R_{ij}\int d^3\bold R_{jk}$. Let $\bold R_{ij}=\bold R_1, \bold R_{jk}=\bold R_2$, and then $R_3=R_{ik}=\sqrt{R_1^2+R_2^2-2R_1R_2\cos\theta}$. Denote by the integral $I=\frac{1}{{\cal N}_r}\int d^3\bold R_{ij}\int d^3\bold R_{jk}
(R_{ij}+R_{jk}-R_{ik})$ with ${\cal N}_r=\int d^3\bold R_{ij}\int d^3\bold R_{jk}$.
To write down the explicit formula of this integral, we apply the restrictions: $R_i\leq R_{i,max}$ and $R_1,R_2\leq R_3$, with $R_{i,max}$ determined through $2aR_{ij}^{max}+\frac{1}{2}\beta(|\epsilon_{i}|+|\epsilon_{j}|+|\epsilon_{i}-\epsilon_{j}|)=\beta\xi_c$ (from the condition $G_{ij}^{min}=G_c$ or $Z_{ij}^{max}=1/G_c$). With the basic triangle geometry (Fig. \ref{triangle}) we obtain
\begin{figure}[ht]
\includegraphics[width=0.3\columnwidth]{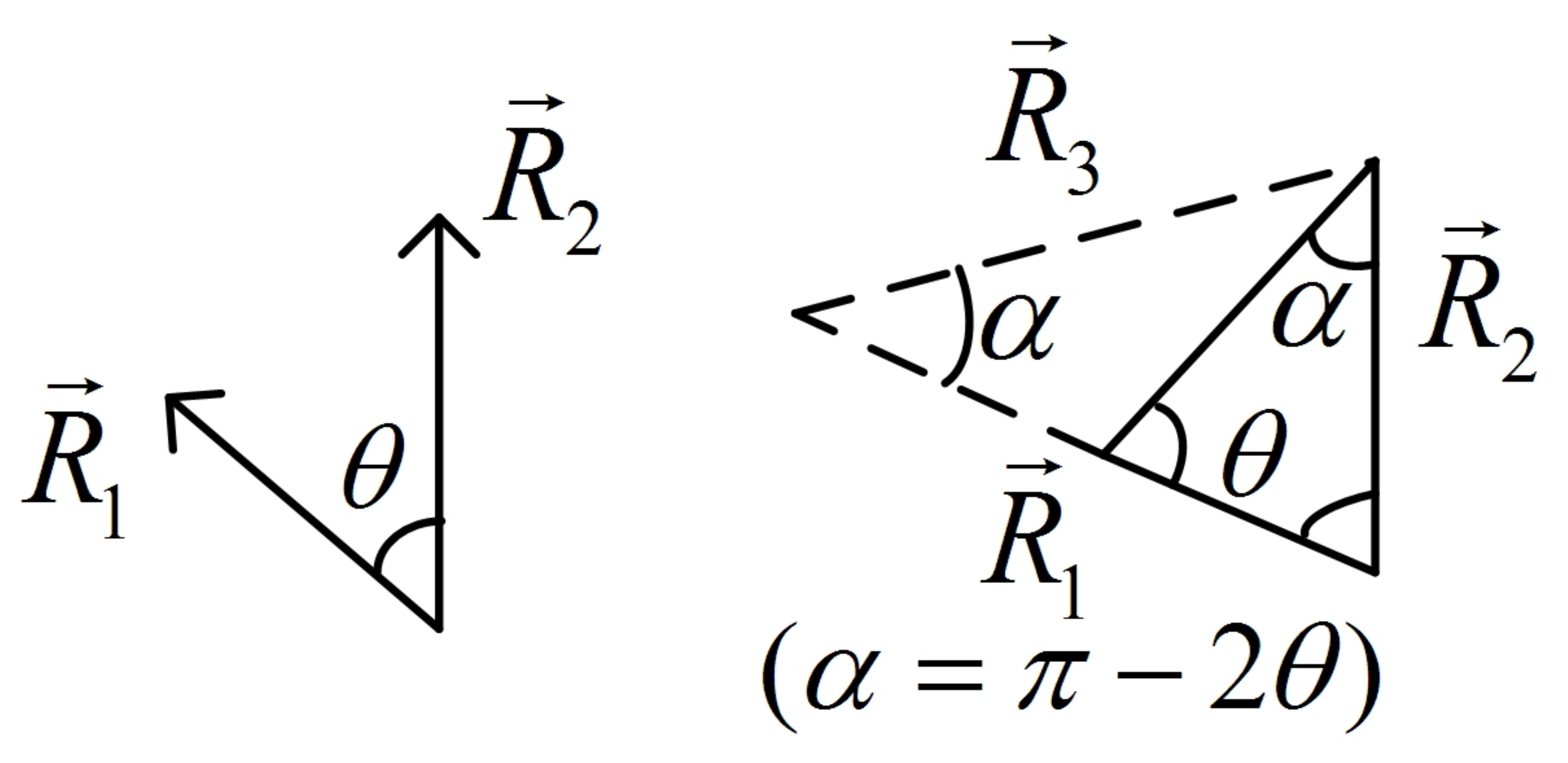}
\caption{(Color online) Triangle geometry for the configuration integral over the position space.}
\label{triangle}
\end{figure}
\begin{eqnarray}\label{eqn:position2}
I&=&\frac{1}{{\cal N}_r}8\pi^2\int_0^{R_{2max}} dR_2R_2^2\bigr[\int_{\pi/2}^{\pi}d\theta\int_0^{R_a}dR_1R_1^2\sin\theta(R_1+R_2-\sqrt{R_1^2+R_2^2-2R_1R_2\cos\theta})\nonumber\\
&&+\int_{\pi/3}^{\pi/2}d\theta\int_{2R_2\cos\theta}^{R_b}dR_1R_1^2\sin\theta(R_1+R_2-\sqrt{R_1^2+R_2^2-2R_1R_2\cos\theta})\bigr],
\end{eqnarray}
where
\begin{eqnarray}\label{eqn:limit1}
R_a&=&\mbox{min}\{R_{1max}, R_2+\sqrt{\Lambda^2-R_2^2\sin^2\theta}\},\nonumber\\
R_b&=&\mbox{min}\{R_{1max}, \frac{R_2}{2\cos\theta}, R_2\cos\theta+\sqrt{\Lambda^2-R_2^2\sin^2\theta}\},
\end{eqnarray}
with
\begin{eqnarray}\label{eqn:limit2}
\Lambda^2=\frac{1}{4a^2}\bigr[\frac{\xi_c}{k_BT}-\frac{1}{2k_BT}(|\varepsilon_{i}|+|\varepsilon_k|+|\varepsilon_i-\varepsilon_k|)\bigr]^2.
\end{eqnarray}
Therefore the integral domain is not uniquely specified and depends on the the integral variables, which makes the Eq. (\ref{eqn:position2}) be still not analytically solvable. We need to simplify it by amplifying the integral domain. From the geometry of the triangle composed of $(R_1,R_2,R_3)$, we can show the following inequalities:
\begin{eqnarray}\label{eqn:inequality1}
\frac{\int_{R_2\cos\theta+\sqrt{\Lambda^2-R_2^2\sin^2\theta}}^{R_{1max}}dR_1R_1^2\sin\theta(R_1+R_2-\sqrt{R_1^2+R_2^2-2R_1R_2\cos\theta})} {\int_{R_2\cos\theta+\sqrt{\Lambda^2-R_2^2\sin^2\theta}}^{R_{1max}}dR_1R_1^2\sin\theta}&\geq&\nonumber\\
\frac{\int_0^{R_2\cos\theta+\sqrt{\Lambda^2-R_2^2\sin^2\theta}}dR_1R_1^2\sin\theta(R_1+R_2-\sqrt{R_1^2+R_2^2-2R_1R_2\cos\theta})} {\int_0^{R_2\cos\theta+\sqrt{\Lambda^2-R_2^2\sin^2\theta}}dR_1R_1^2\sin\theta},
\end{eqnarray}
which is needed in the case $R_2\cos\theta+\sqrt{\Lambda^2-R_2^2\sin^2\theta}<R_{1max}$, and
\begin{eqnarray}\label{eqn:inequality2}
\frac{\int_{R_b}^{R_{1max}}dR_1R_1^2\sin\theta(R_1+R_2-\sqrt{R_1^2+R_2^2-2R_1R_2\cos\theta})} {\int_{R_b}^{R_{1max}}dR_1R_1^2\sin\theta}&\geq&\nonumber\\
\frac{\int_{2R_2\cos\theta}^{R_b}dR_1R_1^2\sin\theta(R_1+R_2-\sqrt{R_1^2+R_2^2-2R_1R_2\cos\theta})} {\int_{2R_2\cos\theta}^{R_b}dR_1R_1^2\sin\theta},
\end{eqnarray}
when $R_b<R_{1max}$. Based on these results, we find that
\begin{eqnarray}\label{eqn:position3}
I&\leq&\frac{1}{{\cal N}_r}8\pi^2\int_0^{R_{2max}} dR_2R_2^2\bigr[\int_{\pi/2}^{\pi}d\theta\int_0^{R_{1max}}dR_1R_1^2\sin\theta(R_1+R_2-\sqrt{R_1^2+R_2^2-2R_1R_2\cos\theta})\nonumber\\
&&+\int_{\pi/3}^{\pi/2}d\theta\int_{2R_2\cos\theta}^{R_{1max}}dR_1R_1^2\sin\theta(R_1+R_2-\sqrt{R_1^2+R_2^2-2R_1R_2\cos\theta})\bigr],
\end{eqnarray}
with
\begin{eqnarray}\label{eqn:normalization2}
{\cal N}_r=8\pi^2\bigr[\int_0^{R_{2max}} dR_2R_2^2\int_{\pi/2}^{\pi}d\theta\int_0^{R_{1max}}dR_1R_1^2\sin\theta+\int_0^{R_{2max}} dR_2R_2^2\int_{\pi/3}^{\pi/2}d\theta\int_{2R_2\cos\theta}^{R_{1max}}dR_1R_1^2\sin\theta\bigr].
\end{eqnarray}
Employing the integral $\int_{\pi/3}^{\pi/2}d\theta\sin\theta\sqrt{R_1^2+R_2^2-2R_1R_2\cos\theta})=\frac{1}{3R_1R_2}
\bigr[(R_1+R_2)^2-(R_1^2+R_2^2-R_1R_2)^{3/2}\bigr]$,
we get finally
\begin{eqnarray}\label{eqn:position5}
I\simeq\frac{2\pi^2}{{\cal N}_r}R_{max}^7-\frac{1.576}{{\cal N}_r}\pi^2R_{max}^7\simeq0.424\pi^2R_{max}^7/{\cal N}_r
\end{eqnarray}
with $R_{max}=\mbox{max}\{R_{1max},R_{2max}\}$. It is easy to obtain the normalization factor as
${\cal N}_r=\frac{23}{18}\pi^2R_{max}^6$.
After the integral over position given above we can now do it over the on-site energies. This gives
\begin{eqnarray}\label{eqn:position9}
\langle R_1+R_2-R_3\rangle_c|_{R_1,R_2<R_3}&=&\frac{0.424}{23/18}\frac{\int d\epsilon_id\epsilon_jd\epsilon_k\rho(\epsilon_i)[n(\epsilon_i)]^3\rho(\epsilon_j)[n(\epsilon_j)]^3 \rho(\epsilon_k)[n(\epsilon_k)]^3
R_{max}^7}{\int d\epsilon_id\epsilon_jd\epsilon_k\int \rho(\epsilon_i)[n(\epsilon_i)]^3\rho(\epsilon_j)[n(\epsilon_j)]^3 \rho(\epsilon_k)[n(\epsilon_k)]^3R_{max}^6}\nonumber\\
&\simeq&0.156\beta\xi_c/a.
\end{eqnarray}
In above calculation we have considered the approximation that the density of states is a constant.

Now we evaluate the average of energy. Similarly, the configurational average of the energy is given by
\begin{eqnarray}\label{eqn:energy1}
&\frac{1}{2k_BT}&\langle |\epsilon_i|+|\epsilon_j|+|\epsilon_j-\epsilon_k|-|\epsilon_i-\epsilon_k| \rangle_c|_{|\epsilon_i|<|\epsilon_j|<|\epsilon_k|}=\nonumber\\
&=&\frac{1}{2k_BT}\frac{\int d\epsilon_id\epsilon_jd\epsilon_k\rho(\epsilon_i)[n(\epsilon_i)]^3\rho(\epsilon_j)[n(\epsilon_j)]^3 \rho(\epsilon_k)[n(\epsilon_k)]^3
(|\epsilon_i|+|\epsilon_j|+|\epsilon_j-\epsilon_k|-|\epsilon_i-\epsilon_k|)}{\int d\epsilon_id\epsilon_jd\epsilon_k
\rho(\epsilon_i)[n(\epsilon_i)]^3\rho(\epsilon_j)[n(\epsilon_j)]^3 \rho(\epsilon_k)[n(\epsilon_k)]^3}.\nonumber\\
\end{eqnarray}
To simplify the above integral, we check $|\epsilon_j-\epsilon_k|-|\epsilon_i-\epsilon_k|$ with the restriction: $|\epsilon_i|<|\epsilon_j|<|\epsilon_k|$. For the case i) $\mbox{Sgn}(\epsilon_i)=\mbox{Sgn}(\epsilon_j)=\mbox{Sgn}(\epsilon_k)=\pm1$, we have
$|\epsilon_j-\epsilon_k|-|\epsilon_i-\epsilon_k|=-|\epsilon_i-\epsilon_j|$;
For ii) $\mbox{Sgn}(\epsilon_i)=\mbox{Sgn}(\epsilon_j)=-\mbox{Sgn}(\epsilon_k)=\pm1$, we have
$|\epsilon_j-\epsilon_k|-|\epsilon_i-\epsilon_k|=-|\epsilon_i-\epsilon_j|$;
For iii) $\mbox{Sgn}(\epsilon_i)=\mbox{Sgn}(\epsilon_k)=-\mbox{Sgn}(\epsilon_j)=\pm1$, we have
$|\epsilon_j-\epsilon_k|-|\epsilon_i-\epsilon_k|=-|\epsilon_i-\epsilon_j|$;
For iv) $\mbox{Sgn}(\epsilon_j)=\mbox{Sgn}(\epsilon_k)=-\mbox{Sgn}(\epsilon_i)=\pm1$, we have
$|\epsilon_j-\epsilon_k|-|\epsilon_i-\epsilon_k|=|\epsilon_i-\epsilon_j|$.
For this we obtain that
\begin{eqnarray}\label{eqn:energy5}
\langle|\epsilon_i|+|\epsilon_j|+|\epsilon_j-\epsilon_k|-|\epsilon_i-\epsilon_k|\rangle_c\simeq \langle|\epsilon_i|+|\epsilon_j|-\frac{1}{2}|\epsilon_i-\epsilon_j|\rangle_c.
\end{eqnarray}
Then by a straightforward calculation one can verify that
\begin{eqnarray}\label{eqn:energy9}
\frac{\beta}{2}\langle |\epsilon_i|+|\epsilon_j|+|\epsilon_j-\epsilon_k|-|\epsilon_i-\epsilon_k| \rangle_c|_{|\epsilon_i|<|\epsilon_j|<|\epsilon_k|}=0.086\beta\xi_c.
\end{eqnarray}
From eqs. (\ref{eqn:position9}) and (\ref{eqn:energy9}) we have
\begin{eqnarray}\label{eqn:total1}
a\langle R_{ij}+R_{jk}-R_{ik}\rangle|_{R_{ij},R_{jk}<R_{ik}}+\frac{\beta}{2}\langle |\epsilon_i|+|\epsilon_j|+|\epsilon_j-\epsilon_k|-|\epsilon_i-\epsilon_k| \rangle_c|_{|\epsilon_i|<|\epsilon_j|<|\epsilon_k|}=0.242\beta\xi_c.
\end{eqnarray}
The lower limit of the AH conductivity is then obtained by
\begin{eqnarray}\label{eqn:minimum4}
\{\sigma_{xy}^{AH}\}_{min}=3L\sigma_{xx}^2\frac{k_BT}{ e^2t_{\substack{max}}^{(0)}}e^{0.242\beta\xi_c}.
\end{eqnarray}
The longitudinal conductivity $\sigma_{xx}$ is given by $G_c$ divided by the correlation length of the network and thus takes the form $\sigma_{xx}=\sigma_0(T)e^{-\beta\xi_c}$ (for the Mott hopping regime, one has $\beta\xi_c=\bigr(\frac{T_0}{T}\bigr)^{1/4}$). We reach further
\begin{eqnarray}\label{eqn:minimum5}
\{\sigma_{xy}^{AH}\}_{min}=3L\sigma_{0}^{0.242}\frac{k_BT}{ e^2t_{\substack{max}}^{(0)}}\sigma_{xx}^{1.758}\propto\sigma_{xx}^\gamma, \ \ \gamma\simeq1.76.
\end{eqnarray}

\section{Upper limit}

Now we show the result of the upper limit, which can be done in a similar procedure. The upper limit is given by
\begin{eqnarray}\label{eqn:anomaloushall2}
\{\sigma_{xy}^{AH}\}_{\substack{max}}\simeq3L\sigma_{xx}^2\frac{k_BT}{ e^2t_{\substack{min}}^{(0)}}\langle R_{ijk}^{\substack{max}}\rangle_c\langle\epsilon_{ijk}^{\substack{max}}\rangle_c,
\end{eqnarray}
where $\langle R_{ijk}^{max}\rangle_c=e^{a\langle R_{ij}+R_{jk}-R_{ik}\rangle_c}|_{R_{ij},R_{jk}>R_{ik}}, \langle\epsilon_{ijk}^{max}\rangle_c=e^{0.5\beta\langle|\epsilon_{i}|+|\epsilon_{j}|+ |\epsilon_{j}-\epsilon_{k}|-|\epsilon_{i}-\epsilon_{k}|\rangle_c}|_{|\epsilon_{i}|>|\epsilon_{j}| >|\epsilon_{k}|}$. To calculate $\langle R_{ij}+R_{jk}-R_{ik}\rangle_c|_{R_{ij},R_{jk}>R_{ik}}$ we again consider first the integral $I=\frac{1}{{\cal N}_r}\int d^3\vec R_1\int d^3\vec R_2
(R_1+R_2-\sqrt{R_1^2+R_2^2-2R_1R_2\cos\theta})$ with ${\cal N}_r=\int d^3\vec R_1\int d^3\vec R_2$.
Note the integral restrictions for the upper limit are: $R_i\leq R_{i,max}$ and $R_1,R_2\geq R_3$, and with the triangle geometry (Fig. (\ref{triangle})) we obtain
\begin{eqnarray}\label{eqn:maxposition2}
I=\frac{1}{{\cal N}_r}8\pi^2\int_0^{R_{2max}} dR_2R_2^2\int_{0}^{\pi/3}d\theta\int_{\frac{R_2}{2\cos\theta}}^{R_a}dR_1R_1^2 \sin\theta(R_1+R_2-\sqrt{R_1^2+R_2^2-2R_1R_2\cos\theta}),\nonumber\\
\end{eqnarray}
where
$R_a=\mbox{min}\{R_{1max}, 2R_2\cos\theta, R_2\cos\theta+\sqrt{\Lambda^2-R_2^2\sin^2\theta}\}$.
Again we simplify the integral by amplifying the integral domain. For this we consider the following inequality:
\begin{eqnarray}\label{eqn:maxinequality1}
\frac{\int_{R_2\cos\theta+\sqrt{\Lambda^2-R_2^2\sin^2\theta}}^{R_{1max}}dR_1R_1^2\sin\theta(R_1+R_2-\sqrt{R_1^2+R_2^2-2R_1R_2\cos\theta})} {\int_{R_2\cos\theta+\sqrt{\Lambda^2-R_2^2\sin^2\theta}}^{R_{1max}}dR_1R_1^2\sin\theta}&\geq&\nonumber\\
\frac{\int_0^{R_2\cos\theta+\sqrt{\Lambda^2-R_2^2\sin^2\theta}}dR_1R_1^2\sin\theta(R_1+R_2-\sqrt{R_1^2+R_2^2-2R_1R_2\cos\theta})} {\int_0^{R_2\cos\theta+\sqrt{\Lambda^2-R_2^2\sin^2\theta}}dR_1R_1^2\sin\theta},
\end{eqnarray}
which is needed in the case $R_2\cos\theta+\sqrt{\Lambda^2-R_2^2\sin^2\theta}<R_{1max}$. With this we find that
\begin{eqnarray}\label{eqn:maxposition3}
I\simeq\frac{1}{{\cal N}_r}8\pi^2\int_0^{R_{2max}} dR_2R_2^2\int_{0}^{\pi/3}d\theta\int_{\frac{R_2}{2\cos\theta}}^{R_{1max}}dR_1R_1^2\sin\theta (R_1+R_2-\sqrt{R_1^2+R_2^2-2R_1R_2\cos\theta}).
\end{eqnarray}
By the same procedure used in the lower limit we obtain $I\simeq0.3729\pi^2R_{max}^7/{\cal N}_r$, and ${\cal N}_r=8\pi^2\int_0^{R_{2max}} dR_2R_2^2\int_{0}^{\pi/3}d\theta\int_{\frac{R_2}{2\cos\theta}}^{R_{1max}}dR_1R_1^2\sin\theta=0.361\pi^2R_{max}^6$.
Further doing the integral over the on-site energies yields
$\langle R_{ij}+R_{jk}-R_{ik}\rangle_c|_{R_{ij},R_{jk}<R_{ik}}=0.483\beta\xi_c/a$.

The configurational average of energy
$\frac{1}{2}\beta\langle |\epsilon_i|+|\epsilon_j|+|\epsilon_i-\epsilon_k|-|\epsilon_j-\epsilon_k| \rangle_c|_{|\epsilon_i|>|\epsilon_j|>|\epsilon_k|}$ can be simplified by checking $|\epsilon_i-\epsilon_k|-|\epsilon_j-\epsilon_k|$ with the restriction: $|\epsilon_i|>|\epsilon_j|>|\epsilon_k|$. Through a similar analysis as applied in the lower limit one can verify $\langle|\epsilon_i|+|\epsilon_j|+|\epsilon_i-\epsilon_k|-|\epsilon_j-\epsilon_k|\rangle_c\simeq \langle|\epsilon_i|+|\epsilon_j|+\frac{1}{2}|\epsilon_i-\epsilon_j|\rangle_c$. Substituting this result into the original integral we obtain finally $\frac{1}{2}\beta\langle |\epsilon_i|+|\epsilon_j|+|\epsilon_j-\epsilon_k|-|\epsilon_i-\epsilon_k| \rangle_c|_{|\epsilon_i|>|\epsilon_j|>|\epsilon_k|}=0.1375\beta\xi_c$.
For this we obtain $\langle R_{ijk}^{max}\rangle_c\simeq e^{0.483\beta\xi_c}$,
$\langle\epsilon_{ijk}^{max}\rangle_c\simeq e^{0.138\beta\xi_c}$, and the upper limit of the AHC by
\begin{eqnarray}\label{eqn:maximum5}
\{\sigma_{xy}^{AH}\}_{max}=3L\sigma_{0}^{0.621}\frac{k_BT}{ e^2t_{\substack{min}}^{(0)}}\sigma_{xx}^{1.379}\propto\sigma_{xx}^\gamma, \ \ \gamma\simeq1.38.
\end{eqnarray}
Based on the results obtained above we thus conclude $\{\sigma_{xy}^{AH}\}\propto\sigma_{xx}^\gamma$ with $1.38<\gamma<1.76$.


\noindent

\end{widetext}


\begin{thebibliography}{99}
\bibitem{AHE1}  N. Nagaosa, J. Sinova, S. Onoda, A. H. MacDonald, and P. Ong, Rev. Mod. Phys. {\bf 82}, 1539 (2010).

\bibitem{Luttinger}  R. Karplus and J. M. Luttinger, Phys. Rev. \textbf{95}, 1154 (1954).

\bibitem{TI} M. Z.  Hasan and C. L. Kane, Rev. Mod. Phys. \textbf{82}, 3045 (2010).

\bibitem{insulating1} B. A. Aronzon et al., 
JETP, {\bf 70} 90 (1999).

\bibitem{insulating2} A. V.  Samoilov etal., 
Phys. Rev. B {\bf 57}, 14032(R) (1998).

\bibitem{insulating3} H. Toyosaki et al., 
Nat. Mater. {\bf 3}, 221 (2004).

\bibitem{insulating4} W. Allen et al., 
Phys. Rev. B \textbf{70}, 125320 (2004).

\bibitem{insulating5} Sh. U. Yuldashev et al., Phys. Rev. B \textbf{70}, 193203 (2004).

\bibitem{insulating6} K. Ueno et al., 
Appl. Phys. Lett. \textbf{90}, 072103 (2007).

\bibitem{insulating7} S. Shen, et al., J. Appl. Phys. {\bf 103}, 07D134 (2008).

\bibitem{insulating8} A. Fern\'{a}ndez-Pacheco, et al., Phys. Rev. B \textbf{77}, 100403(R) (2008).

\bibitem{insulating9} D. Venkateshvaran, et al., Phys. Rev. B \textbf{78}, 092405 (2008).

\bibitem{insulating10} M. Glunk et al., 
Phys. Rev. B \textbf{80}, 125204 (2009).

\bibitem{insulating11} D. Chiba et al., Phys. Rev. Lett. \textbf{104}, 106601 (2010).
\bibitem{Onoda} S. Onoda, N. Sugimoto, and N. Nagaosa, Phys. Rev. Lett. {\bf 97}, 126602 (2006).
\bibitem{Manganites} S. H. Chun et al., 
Phys. Rev. Lett. {\bf 84}, 757 (2000).

\bibitem{Burkov} A. A. Burkov and L. Balents, Phys. Rev. Lett. {\bf 91}, 057202 (2003).

\bibitem{note} A metallic theory introducing strong disorder broadening showed an above unity scaling outside its range of validity ($k_F l \ll 1$),
but predicts, expectedly, metallic conductivities at zero temperature and is therefore invalid in the insulating regime \cite{Onoda,AHE1}.

\bibitem{Abrahams}  A. Miller and E. Abrahams, Phys. Rev. {\bf 120}, 745 (1960).

\bibitem{Mott} N. F. Mott, Phil. Mag. \textbf{19}, 835 (1969).

\bibitem{Holstein} T. Holstein, Phys. Rev. {\bf 1}, 1329 (1961).

\bibitem{Halperin} V. Ambegaokar al., 
Phys. Rev. B {\bf 4}, 2612 (1971).

\bibitem{Pollak} M. Pollak, J. Non-Cryst. Solids \textbf{11}, 1-24 (1972).


\bibitem{hopping1} G. E. Pike et al., 
Phys. Rev. B \textbf{10}, 1421 (1974).

\bibitem{hopping2} H. Overhof, Phys. Stat. Sol. (b) \textbf{67}, 709 (1975).

\bibitem{DMS} G. A. Fiete et al., 
Phys. Rev. Lett. \textbf{91}, 097202 (2003).

\bibitem{Efros} A. L. Efros and B. I. Shklovskii, J. Phys. C {\bf 8}, L49 (1975).


\end{thebibliography}

\begin{thebibliography}{99}
\bibitem{DMS} G. A. Fiete, G. Zar\'{a}nd, and K. Damle, 
Phys. Rev. Lett. \textbf{91}, 097202 (2003).

\bibitem{Pollak} M. Pollak, 
J. Non-Cryst. Solids \textbf{11}, 1-24 (1972).

\bibitem{E-S} A. L. Efros and B. I. Shklovskii, J. Phys. C {\bf 8}, L49 (1975).

\bibitem{perturbation} H. B\"{o}ttger and V. V. Bryksin., 
phys. stat. sol. (b) \textbf{81}, 433 (1977).
\end{thebibliography}
\end{document}